\documentclass[twocolumn,prl,showpacs,superscriptaddress]{revtex4}
\usepackage{amsmath}
\usepackage{amssymb}
\usepackage{bm}
\usepackage{ulem}
\usepackage{epsfig}
\usepackage{graphicx}
\usepackage{xcolor}

\begin{document}

\title{Resonant Fibonacci Quantum Well Structures}

\author{A.N.~Poddubny}\email{poddubny@coherent.ioffe.ru}
\affiliation{A.F.~Ioffe Physico-Technical Institute, 194021
St.~Petersburg, Russia}
\author{L.~Pilozzi}
\affiliation{Istituto dei Sistemi Complessi,
CNR, C. P. 10, Monterotondo Stazione, Rome I-00016,Italy}
\author{M.M.~Voronov}
\author {E.L.~Ivchenko}
\affiliation{A.F.~Ioffe Physico-Technical Institute, 194021
St.~Petersburg, Russia}

\begin{abstract}
 We propose a resonant one-dimensional quasicrystal, namely,
a multiple quantum well (MQW) structure satisfying the
Fibonacci-chain rule with the golden ratio between the long and
short inter-well distances. The resonant Bragg condition is generalized
from the periodic to Fibonacci MQWs. A dispersion equation for
exciton-polaritons is derived in the two-wave approximation, the
effective allowed and forbidden bands are found. The reflection
spectra from the proposed structures are calculated as a function
of the well number and detuning
from the Bragg condition.
\end{abstract}
\maketitle
\section{Introduction}
The concept of quasicrystal as a non-periodic structure with
perfect long-ranged bond orientational order was brought in
solid-state physics by Levine and Steinhardt~\cite{levine}. It was
extended to optics in Ref.~[\onlinecite{kohmoto}], where a
one-dimensional (1D) quasicrystal model constructed of dielectric
layers forming the Fibonacci sequence was proposed. At just the same
time the concept of photonic crystals was suggested by
Yablonovich~\cite{yablonovich} and John~\cite{john}. Since then
the 1D photonic Fibonacci quasicrystals have been extensively
studied\cite{hattori,kalit,peng,zhukovsky}.

In this paper we introduce a new nanoobject, the Fibonacci quantum
well (QW) structure with inter-well spacings arranged in the
Fibonacci sequence. This means that the thickness of barriers
separating the wells can take one of two values so that the ratio
between the long and short inter-well spacings equals the golden
mean $\tau = (\sqrt{5}+1)/2$. We focus on the light propagation in
such a medium in the frequency region around the resonance
frequency $\omega_0$ of a two-dimensional exciton in the quantum
well. The barriers are assumed to be thick enough so that the
excitons in different wells are coupled only via electromagnetic
field. Thus, the object under study is a resonant photonic
quasicrystal, an intermediate structure between completely ordered
and disordered media, namely, periodic MQWs with a fixed
inter-well spacing and MQWs with random inter-well spacing.

Among periodic QW structures, of particular interest are the
resonant Bragg structures with the period satisfying the Bragg
condition
\begin{equation}\label{bragg}
q(\omega_0) d = \pi j,\: j=1,2\ldots
\end{equation}
where $q(\omega)=\omega n_b/c$ and $q(\omega_0)$ is the light wave
vector at the exciton resonance frequency $\omega_0$, $n_b$ is the
background refractive index of both QW and barrier materials, $d$ is the
structure period, and $c$ is the light velocity. The periodic
resonant Bragg MQWs have been first considered theoretically in
Ref.~[\onlinecite{ftt}] and then investigated in a number of
theoretical as well as experimental works
\cite{kocher,merle1,prineas,IvchenkoWillander,Ikawa_Cho,voronov,pilozzi,pilozzi2007}.
It was established that, for small enough numbers $N$ of QWs
(superradiant regime), the optical reflection spectrum is
described by a Lorentzian with the halfwidth $N \Gamma_0 +
\Gamma$, where $\Gamma_0$ and $\Gamma$ are, respectively, the
exciton radiative and nonradiative damping rates in a single
QW~\cite{ftt,Ikawa_Cho}. For a large number of wells (photonic
crystal regime), the reflection coefficient is close to unity
within the forbidden gap for exciton polaritons propagating in infinite
periodic system and rapidly decreases near the gap edges $\omega_0
- \Delta/\sqrt{j}$ and $\omega_0 + \Delta/\sqrt{j}$, where $\Delta
= \sqrt{2 \omega_0 \Gamma_0/\pi
}$~~\cite{IvchenkoWillander,Ikawa_Cho,pilozzi,voronov}.

In Section II we will show that a generalized resonant Bragg
condition analogous to Eq.~(\ref{bragg}) can be formulated for
the resonant Fibonacci MQW structures, although the latter are
aperiodic. In Section III the significance of the proposed
condition is verified by numerical calculations of the reflection
spectra from the structures tuned on and slightly detuned from this
condition, and the dependence of the reflection spectra on the
number of wells is analyzed and compared with those for the
periodic Bragg structures. In Section IV we apply a two-wave
approximation in order to determine the band gaps in the exciton-polariton  spectrum of the Fibonacci structures and show that the 
simple analytic theory allows one to interpret quite well the
numerical results.
\begin{figure}[b]
\begin{center}
\includegraphics[width=0.45\textwidth]{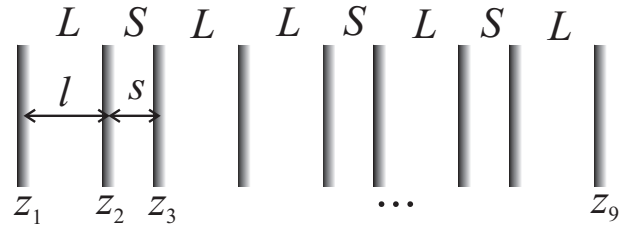}
\end{center}
\caption{Scheme of the Fibonacci QW structure ${\cal F}_6$ with $N=9$ QWs.\label{f1}}
\end{figure} 

\section{Resonant Bragg condition for Fibonacci MQW structure}
The structure under consideration is schematically depicted in
Fig.~1. It consists of $N$ identical QWs embedded in a matrix with
dielectric constant $\varepsilon_{b}$.
 The inter-well distances
take two values represented by long and short segments of 
length $l$ and $s$, respectively. For the Fibonacci chain, the
coordinate $z_m$ of the $m$-th QW center is given by\cite{janot}
\begin{equation}\label{xj}
z_m=\bar d (m-1) + \frac{s}{\tau}\left(\frac{1}{\tau}-\left\{\frac{m}{\tau}\right\}\right),
\end{equation}
where the integer $m$ runs from 1 to $N$, $\tau$ is the golden
ratio, $\bar d$ is the average period of the structure given by the
product $s(3-\tau)$, and $\{ x \}$ is the fractional part of $x$.
An alternative way of defining $z_m$ is based on the recurrence
relation ${\mathcal F}_{j+1}= \{{\mathcal F}_{j}, {\mathcal
F}_{j-1}\}$ for finite Fibonacci chains of the order $j + 1$, $j$
and $j - 1$, with initial conditions ${\mathcal F}_{1}={S}$,
${\mathcal F}_{2}={L}$, where $S$ and $L$ are the segments with
lengths $s$ and $l=\tau s$, respectively.\cite{janot} Then $z_m$
are coordinates of boundaries between the segments in the
${\mathcal F}{_j}$ sequence.

The exact reflection coefficient of the light normally incident on
such a structure from the left half-space can be obtained by
standard transfer matrix method \cite{book}. In order to form the
base for formulation of the resonant Bragg condition for the
Fibonacci structures we will analyze the reflection in the
first-order Born approximation neglecting multireflection
processes and summing up the amplitudes of waves reflected from
distinct wells. Then the amplitude reflection coefficient,
$r_N(\omega)$, from the $N$-well Fibonacci structure at the light
frequency $\omega$ is given by
\begin{equation} \label{B1}
r_N(\omega) \approx N\; f[q(\omega),N]\; r_1(\omega)\:,
\end{equation}
where $f(q,N)$ is the structure factor of the system,
\[
f(q,N) = \frac{1}{N} \sum_{m = 1}^{N}{\rm e}^{2 {\rm i} q z_m}\:,
\]
and $r_1$ is the reflection coefficient from a single QW,
\[ r_1(\omega) = \frac{{\rm i} \Gamma_0}{\omega_0 - \omega - {\rm i}
(\Gamma + \Gamma_0)}\:.
\]
For the semiinfinite Fibonacci MQWs the structure factor $f(q) =
\lim_{N \to \infty} f(q,N)$ can be presented in the following
analytical form\cite{janot}
\begin{gather}\label{ffc1}
f(q) = \sum\limits_{h,h' = - \infty}^{\infty} \delta_{2q,G_{hh'}}
f_{hh'},\quad G_{hh'}
= \frac{2\pi}{\bar{d}} (h + h'/\tau),\\
f_{hh'} = \frac{\sin S_{hh'}}{S_{hh'}} \exp\Bigl({\rm i}
\frac{\tau-2}{\tau} S_{hh'}\Bigr),\label{ffc2} \\ \quad S_{hh'} =\frac{\pi
\tau}{\tau^2+1}(\tau h'-h).\nonumber
\end{gather}
Allowed diffraction vectors $G_{hh'}$ form a dense
pseudocontinuous set. The largest values of $|f_{hh'}|$ are
reached for the pairs $(h,h')$ coinciding with two successive
Fibonacci numbers ($F_{j}$, $F_{j-1}$) with $F_j$ defined recursively by $F_0
= 0, F_1 = 1$ and $F_{j+1} = F_{j} + F_{j-1}$. Thus, for $(h,h') =
(F_{j}, F_{j-1}) = (1,0), (1,1), (2,1), (3,2)$ and (5,3)
corresponding to $j = 1 ... 5$, the modulus of $f_{hh'}$ equals to
$\approx$ 0.70, 0.88, 0.95, 0.98 and 0.99, respectively. For
$(h,h')$ not belonging to this particular set, values of
$|f_{hh'}|$ are  significantly smaller. It follows then that if the exciton
resonance frequency satisfies the condition
\begin{equation}\label{bragg2}
\frac{\omega_0 n_b}{c}\bar d = \pi \left( F_{j} + \frac{F_{j-1}}{\tau} \right) ,\quad j=1,2\ldots
\end{equation}
the coefficient (\ref{B1}) at $\omega = \omega_0$ and large $N$
amounts to
\[
r_N = N f[q(\omega_0),N] r_1(\omega_0) \approx - \frac{N\Gamma_0 f_{hh'} }{\Gamma_0 + \Gamma}\:.
\]
This is of the same order of magnitude as the reflection
coefficient calculated in the same Born approximation for a
periodic resonant Bragg structure satisfying Eq.~(\ref{bragg}).
Hence  Eq. (\ref{bragg2}) is indeed a resonant Bragg
condition generalized for the Fibonacci MQWs. In the following we
fix the value of $\omega_0$, consider the average period ${\bar
d}$ as a variable parameter and use the notation ${\bar d}_j$ for
${\bar d}$ given by Eq.~(\ref{bragg2}) for the integer $j$. The
corresponding thicknesses $s_j, l_j$ of the short and long
segments are related with ${\bar d}_j$ by
\begin{equation}
s_j = {\bar d}_j/(3 - \tau)\:,\: l_j = {\bar d}_j \tau /(3 - \tau)\:.
\end{equation}

The estimation (\ref{B1}) for $r_N$ is valid until $|r_N| \ll 1$,
i.e., if $N \Gamma_0 \ll {\rm max} \{ |\omega_0 - \omega|, \Gamma
\}$. Otherwise one has to take into account the multireflection of
the light waves from QWs which is readily achieved by the standard
transfer-matrix numerical calculation. The results are presented
and analyzed in the next section.

\section{Calculated reflection spectra}
\begin{figure}[t]
\begin{center}
\includegraphics[width=0.45\textwidth]{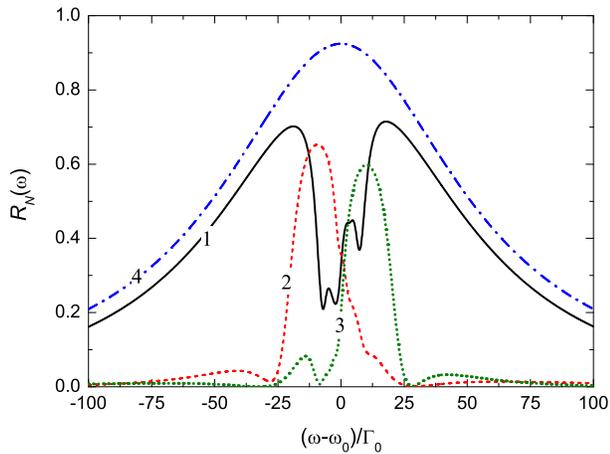}
\end{center}
\caption{Reflection spectra calculated for three
Fibonacci structures satisfying the resonant Bragg condition \eqref{bragg2} for $j = 2$
(curve 1) and detuned by $\pm 2\%$ from this condition (curves 2 and 3) in comparison with the reflection spectrum
from the periodic resonant
Bragg QW structure (curve 4). The values of parameters are indicated in text. \label{f2}
}
\end{figure}

Figure 2 presents reflection spectra calculated for four
structures containing $N=50$ quantum wells. The exciton parameters
used are as follows: $\hbar \omega_0 = 1.533$~eV, $\hbar\Gamma_0 =
50~\mu$eV, $\hbar\Gamma = 100~\mu$eV, $n_b=3.55$. Curve 1 is calculated
for the resonant Fibonacci QW structure satisfying the exact Bragg
condition (\ref{bragg2}) with $j=2$ so that ${\bar d} = {\bar
d}_2$, $s = s_2$ and $l = l_2$. Curves 2 and 3 correspond to the
Fibonacci structures with the barrier thicknesses slightly detuned
from $s_2$ and $l_2$: $s/s_2 = l/l_2 = 1.02$ for curve 2 and
$s/s_2 = l/l_2 = 0.98$ for curve 3. Curve 4 describes the
reflection from the periodic Bragg structure with the same
exciton parameters and the period $d = \pi/q(\omega_0)$, satisfying
Eq.~(\ref{bragg}). From comparison of curves 1 and 4 we conclude
that the reflection spectra from the resonant periodic and
Fibonacci structures tuned to the Bragg conditions (\ref{bragg})
and (\ref{bragg2}) are close to each other outside the frequency
region around $\omega_0$.  Moreover it follows from curves 2 and 3 that a
slight deviation from the condition (\ref{bragg2}) results in a
radical decrease of the effective spectral halfwidth. This sensitivity is a characteristic feature of Fibonacci structures, as it happens for the periodic Bragg QW systems. The remarkable structured dip in the middle of the
spectrum 1 is the only qualitative difference from the periodic
structures, the origin of this dip is explained in the next
section. Now we turn to analysis of reflection spectra as a
function of the QW number $N$ and index $j$ in
Eq.~(\ref{bragg2}).

Evolution of the reflection spectra with the QW number $N$ is
illustrated in Fig.~3a. The spectral envelope smoothed to ignore
 dip in the middle shows a behavior similar to that of the
conventional Bragg QW structure. Indeed, for small $N$ the
envelope is a Lorentzian with the halfwidth increasing as a linear
function of $N$. This is a straightforward manifestation of
superradiant regime, which, as we can see here, does not necessarily
require periodicity even if the inter-well distances are
comparable to the light wavelength. The saturation of the
spectral halfwidth (photonic crystal regime) begins at large $N$
of the order of $ \sqrt{\omega_0/\Gamma_0}$, in a similar way as for the periodic Bragg structures.  The shape of the
spectra for large $N$ allows us to suppose existence of two
wide symmetrical stop bands in the energy spectrum of the
structure with an allowed band between them. Of course, the
application of terms ``allowed'' and ``stop'' bands to an aperiodic
structure is questionable. In section IV we show that nevertheless
these terms are applicable in a reasonable approximation.

Figure 3b presents the reflection spectra of the Fibonacci QW structures
containing a large number of wells, $N = 200$, and satisfying
Eq.~\eqref{bragg2} with three different values of $j$. All the
curves indicate an existence of the stop and allowed bands.
However, the band widths are $j$-dependent: the stop band (or gap)
indicated by a united pair of vertical lines and the middle dip
are both squeezed with increase of $j$.

\begin{figure}[t]
\begin{center}
\includegraphics[width=0.5\textwidth]{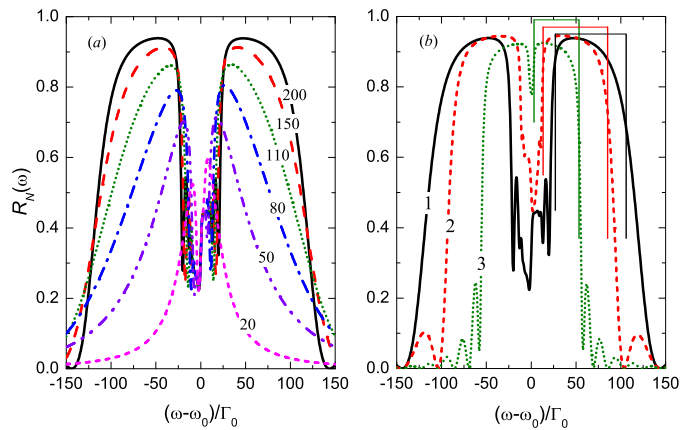}
\end{center}\caption{Reflection spectra from the resonant Fibonacci structures.
(a) Six curves are calculated for the structures satisfying the condition (\ref{bragg2}) with $j=2$
and $N = 20, 50, 80, 110, 150, 200$. The number of wells is indicated near each corresponding curve.
(b) Curves 1, 2, 3 are calculated for the structures with $N=200$ and  indices $j = 2, 3, 5$ in Eq.~\eqref{bragg2}.
Vertical lines connected by a horizontal bar indicate
the exciton-polariton high-frequency gap given by Eq.~\eqref{ws}. Other parameters are the same as in Fig.~2.
}
\end{figure}
\section{Exciton-polariton energy spectrum}
For the light propagating in a system of identical QWs located at
the points $z_m$ ($m = 1, 2...$), the equation for the electric field can be written as \cite{book}
\begin{equation}\label{m1}\begin{split}
\left(-\frac{d^2}{dz^2}-q^2\right)& E(z) =\\ &\frac{2 q \Gamma_0}{\omega_0 -
\omega - {\rm i} \Gamma}\sum_{m}\delta(z - z_m)E(z_m)\:,\end{split}
\end{equation}
where $q \equiv q(\omega)$ and we assume that  quantum wells
are thin as compared to the light wavelength. We consider
a semiinfinite Fibonacci QW structure with the average period
${\bar d}_j$ satisfying Eq.~(\ref{bragg2}) for a certain value of
$j$. Using the above mentioned properties of the
coefficients $f_{hh'}$ in the structure factor $f(q)$ we can
retain in the sum (\ref{ffc1}) only one term $f_{hh'} \delta_{2 q, G_{hh'}}$ with $(h,h') = (F_j, F_{j-1})$. In other words we take
into account only one diffraction vector $G_{hh'}$ corresponding
to the condition \eqref{bragg2} and neglect all other possible
diffraction vectors. In this approximation we can present the
light wave as a sum of two plane waves with the wave vectors $K$
and $K' = K - G_{hh'}$ assuming $K \approx G_{hh'}/2$. The
amplitudes of the chosen spatial harmonics, $E_K$ and $E_{K'}$,
satisfy the following two coupled equations
\begin{eqnarray}\label{w2}
&&(q^2-K^2+\chi) E_K  +  \chi f^{*}_{hh'} E_{K'} = 0\:, \\
&&\chi f_{hh'} E_{K} + (q^2-K'^2 +   \chi) E_{K'} = 0\:, \nonumber
\end{eqnarray}
where
\[
\chi = \frac{2q\Gamma_0}{\bar d(\omega_0 - \omega - {\rm i} \Gamma)}\:.
\]

In the following analysis we ignore the exciton dissipation, neglecting the nonradiative damping. Thus the frequency axis is divided into intervals of purely allowed and forbidden
bands with propagating and evanescent polaritonic solutions. In
the allowed bands the solutions are characterized by real values
of the wave vector $K$. It is convenient to reduce the
exciton-polariton dispersion $\omega(K)$ to the ``first Brillouin
zone'' defined in the interval $- G_{hh'}/2 < K \leq G_{hh'}/2$.
The detailed behavior of $\omega(K)$ inside this interval lies out
of the scope of the present paper. Note that, in close vicinity to
$\omega_0$, the two-wave approximation is inadequate and the
polariton dispersion should be calculated taking into account an
admixture of a lot of plane waves. Here we consider only the
exciton-polariton eigenfrequencies at the edge of the Brillouin
zone, $K = - K' = G_{hh'}/2$. It follows from Eq.~(\ref{w2}) that
four eigenfrequencies at this point are given by
\begin{gather}\label{ws}
\omega_{\rm out}^{\pm}=\omega_0\pm \Delta \sqrt{\frac{1 + |f_{hh'}|}{ 2\:(h+h'/\tau)}}\:,\\
\omega_{\rm in}^{\pm}=\omega_0\pm \Delta \sqrt{\frac{1 -
|f_{hh'}|}{ 2\:(h + h'/\tau)}} \nonumber \:.
\end{gather}
In accordance with Fig.~3 we attribute the interval $\omega_{\rm
in}^+ < \omega < \omega_{\rm out}^+$ to the exciton-polariton
upper stop band (labelled by  index ``+'') and the interval
between $\omega_{\rm out}^-$ and $\omega_{\rm in}^-$ to the lower
stop band (labelled by index ``--''). The subscripts  ``in'' and
``out'' denote the stop-band edges inner and outer with respect to
$\omega_0$.

The values of $\omega_{\rm in}^+$ and $\omega_{\rm out}^+$ are
marked by vertical lines in Fig.~3b. One can see an excellent
agreement between the band edges revealed in the calculated
spectra and those given by Eq.~\eqref{ws} which unambiguously
confirms the interpretation of the frequencies (\ref{ws}).

Equations (\ref{ws}) can be reduced to those for the periodic resonant
Bragg structures as soon as $|f_{hh'}|$ is set to unity and $h +
h'/\tau$ is replaced by the integer $j$. For
$|f_{hh'}| = 1$ the inner eigenfrequencies merge at $\omega_0$  and a single band gap of width $2 \Delta/\sqrt{j}$ is formed. In the
Fibonacci QW structures $|f_{hh'}| < 1$ and, as a result, an
allowed band opens between $\omega_{\rm in}^-$ and $\omega_{\rm
in}^+$. We note that a qualitatively similar band structure can
be realized when the periodic MQWs has a compound elementary cell.\cite{voronov}
One can easily show that, also in this case, the modulus of the
structure factor is smaller than unity. Moreover,
Eqs.~\eqref{ws} can be reduced to Eqs.~(26) of Ref.
[\onlinecite{voronov}] if $h + h'/\tau$ and $|f_{hh'}|$ are
replaced, respectively, by 1/2 and $|\cos{q d_2}|$, where $d_2$ is
the inter-well distance in the compound unit cell of a periodic
structure with two QWs in the supercell.

In the Fibonacci QW structure the decrease of stop-band widths with the increasing $h +
h'/\tau$ is related to the corresponding increase of the average
period ${\bar d}$ in Eq.~(\ref{bragg2}) and it is analogous to
the $j^{-1/2}$ power law of the band width for the periodic
resonant Bragg structures. The middle allowed band width decreases
even faster because, as mentioned above, the value of $|f_{hh'}|$
tends to unity and, therefore, the value of $\sqrt{1 - |f_{hh'}|}$
rapidly vanishes as the index $j$ in Eq.~(\ref{bragg2}) changes
from 2 to 5.
\section{Conclusions}
We have introduced into consideration resonant 1D photonic
quasicrystals based on Fibonacci QW structures. The analysis of
light reflection in the Born approximation has been used to
formulate the resonant Bragg condition for this system. The results of  straightforward transfer-matrix
numerical calculation confirm the relevance of the generalized
Bragg condition imposed on the aperiodic system under study. For a
small number $N$ of QWs, the Fibonacci structures show the
superradiant behavior while, for high values of $N$ exceeding
$\sqrt{\omega_0/\Gamma_0}$, the photonic crystal
regime with distinct stop bands in optical spectra is reached. A qualitative
difference with respect to the periodic resonant Bragg QW structures lies in
the presence of a structured dip in the reflection spectrum around
the exciton resonance frequency $\omega_0$. An approximate
two-wave exciton-polariton model allows one to describe the widths
of the allowed and forbidden bands as a function of the structure
parameters.
\acknowledgements {This work was supported by RBFR and the ``Dynasty'' Foundation -- ICFPM.}

\end{document}